\documentclass[conference]{IEEEtran}
\IEEEoverridecommandlockouts
\usepackage{cite}
\usepackage{amsmath,amssymb,amsfonts}
\usepackage{algorithmic}
\usepackage{graphicx}
\usepackage{textcomp}
\usepackage{subcaption}
\usepackage{xcolor}
\usepackage{multirow,tabularx}
\usepackage{booktabs}
\usepackage[normalem]{ulem}
\def\BibTeX{{\rm B\kern-.05em{\sc i\kern-.025em b}\kern-.08em
    T\kern-.1667em\lower.7ex\hbox{E}\kern-.125emX}}
\makeatletter

\newcommand\blfootnote[1]{%
  \begingroup
  \renewcommand\thefootnote{}\footnote{#1}%
  \addtocounter{footnote}{-1}%
  \endgroup
}

\begin{document}

\title{CONTENT-ADAPTIVE MOTION RATE ADAPTION FOR LEARNED VIDEO COMPRESSION}
\author{
    % \vspace{-0.5em}
    \begin{tabular}{cccc}
        Chih-Hsuan Lin & 
        Yi-Hsin Chen &
        Wen-Hsiao Peng &
    \end{tabular}\\
    
    {\tt\small {\{meerkat10.cs09g@, yhchen.iie07g@, wpeng@cs.\}nctu.edu.tw}}\\[1ex]
    
    Computer Science Dept., National Yang Ming Chiao Tung University, Taiwan
    \vspace{-3ex}
}
% \author{
%     \begin{tabular}{cccc}
%         Yung-Han Ho$^{1}$ &
%         Chih-Hsuan Lin$^{1}$ &
%         Peng-Yu Chen$^{1}$ & 
%         Mu-Jung Chen$^{1}$\\
%     \end{tabular}\\
%     %\vspace{1em}
%     \begin{tabular}{ccc}
%         Chih-Peng Chang$^{1}$ &
%         Wen-Hsiao Peng$^{1}$ &
%         Hsueh-Ming Hang$^{2}$\\%[1ex]
%     \end{tabular}\\
%     %{\tt\small {\{hectorho0409.cs04g@, dororojames.cs07g, wpeng@cs., cpchang.cs08g@\}nctu.edu.tw}}\\[1ex]
    
%     $^1$Computer Science Dept., 
%     $^2$Electronics Engineering Dept., 
%      National Yang Ming Chiao Tung University, Taiwan
%     \vspace{-2ex}
% }

%\author{A. Jones\\University of Somewhere \and
%B. Smith\\University of Somewhere Else}

\maketitle

\begin{abstract}
This paper introduces an online motion rate adaptation scheme for learned video compression, with the aim of achieving content-adaptive coding on individual test sequences to mitigate the domain gap between training and test data. It features a patch-level bit allocation map, termed the \textit{$\alpha$-map}, to trade off between the bit rates for motion and inter-frame coding in a spatially-adaptive manner. We optimize the \textit{$\alpha$-map} through an online back-propagation scheme at inference time. Moreover, we incorporate a look-ahead mechanism to consider its impact on future frames. Extensive experimental results confirm that the proposed scheme, when integrated into a conditional learned video codec, is able to adapt motion bit rate effectively, showing much improved rate-distortion performance particularly on test sequences with complicated motion characteristics. 

\end{abstract}

\blfootnote{This work was supported by the Higher Education Sprout Project of the National Yang Ming Chiao Tung University, Ministry of Education (MOE), Taiwan, MOST, Taiwan, under Grant MOST 110-2221-E-A49-065-MY3 and National Center for High-performance Computing.}
\section{Introduction}
End-to-end learned video compression has been an active research area since the advent of the seminal work DVC~\cite{dvclu}. Much research has been focused on improving temporal prediction for residual coding in the pixel domain~\cite{dvcpro, mlvc, elfvc,ssf} or the feature domain~\cite{fvc}. Recently, a new school of thoughts emerged, replacing residual coding with conditional coding~\cite{mmsp,dcvc,canf} and making a significant breakthrough in compression performance. 

Although showing promising coding results, learned video codecs may suffer from generalization issues. The domain gap between training and test data often leads to their sub-optimal coding performance and/or poor generation on individual test sequences.  

%Recently, points out the inefficiency of the residual coding and introduces the conditional coding. Follow this line, \cite{dcvc} introduces more informative condition for conditional codec and proposes a conditional temporal prior to further improve the entropy coding. \cite{canf} proposes a purely conditional coding framework, which contains conditional inter-frame and motion codec.

%It is noted in~\cite{ling2022future} that the traditional video encoding process usually involves time-consuming rate-distortion optimization, with the aim of adapting coding modes to every video frame. Thus, the encoding complexity can be orders of magnitude higher than the decoding complexity. In contrast, learned video codecs are less adaptable to changes in input statistics, generally showing more balanced encoding and decoding complexity. 

%in order to 

With the aim of achieving better rate-distortion performance on individual test sequences at inference time, content-adaptive coding with learned codecs attracts lots of attention. \cite{campos2019content, zou20202, lu2020content} propose to optimize the latent representations of individual images/sequences or the encoder through back-propagating a rate-distortion loss at inference time. \cite{van2021instance,zhang2021learn} further include the decoder for end-to-end optimization and signal the optimal decoder parameters in the bitstream. Taking a different approach,~\cite{brand2021rate} reserves external coding options for learned codecs to perform resolution-adaptive coding of the flow map or the image latents. Similarly,~\cite{qmap} creates a pixel-wise importance map that can be specified for spatially-adaptive image coding. While the aforementioned methods may invoke back-propagation or rate-distortion optimization at inference time,~\cite{hu2022coarse} represents a feed-forward approach that uses a neural network to predict coding modes from observing the hyperprior for resolution-adaptive flow and residual coding.

%In a sense, it is similar to variable block size motion compensation and residual coding.

%Learned video compression has been grow strongly in computer vision tasks since DVC~\cite{dvclu}, which brings the concept of predictive coding in traditional codecs. Afterwards, several works including~\cite{dvcpro, mlvc, elfvc, fvc} keep working on end-to-end learned video compression based on residual coding. Recently,~\cite{mmsp} pointed out the inefficiency of the residual coding and introduce the conditional coding. \cite{dcvc} improve this work by introducing stronger condition for conditional coder, and a conditional temporal prior is proposed to further improve the entropy coding. Furthermore, \cite{canf} proposed the augmented normalizing flow-based conditional inter-frame coder and introduce this idea to motion coder. However, learned codecs usually have generalization issue, which may perform poorly on unseen data. Also, it is shown in [16] that the traditional image/video encoding process takes a much longer run-time than the decoding process, while the learned image/video codecs have similer encoding and decoding run-times. This is attributed to the rate-distortion optimization process in the traditional encoder, which aims to adapt coding modes to every video frame.

Inspired by~\cite{qmap}, we introduce a patch-level bit allocation map, termed the \textit{$\alpha$-map}, to a learned video codec. The \textit{$\alpha$-map} offers a mechanism to trade off between the bit rates for motion and inter-frame coding in a spatially-adaptive manner. To optimize the \textit{$\alpha$-map} for content-adaptive coding, we propose an online back-propagation scheme with look-ahead to consider its impact on future frames. Extensive experimental results confirm the effectiveness of the proposed method.

%By adjusting the \textit{$\alpha$-map}, our framework is able to spatially-varying trade-off the bit-allocation between motion and inter-frame codec for individual test sequences. 
%To address this issue, we propose to introduce an external signal for learned video compression that can be utilized for content-adaptation. By adjusting the external signal, which is actually modifying the bit-allocation between motion and inter-frame coder. Furthermore, we introduce two efficient algorithms to update the external signal through back-propagation in inference time. The experimental results show the superiority of our methods.
\vspace{-.3em}
\section{Related Work}
% \vspace{-1em}
%\subsection{Image Compression Using Augmented Normalizing Flows}
\label{sec:canf}

\begin{figure}
% \vspace{-.6em}
\vspace{-0.5cm}
\centering
\includegraphics[width=0.6\linewidth]{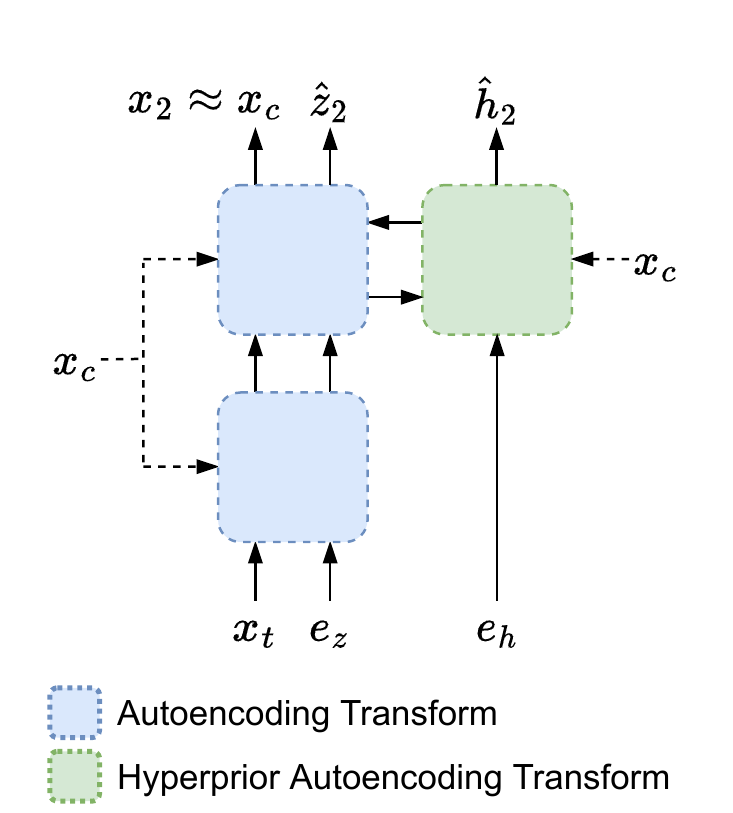}
\caption{Conditional augmented normalizing flow-based inter-frame codec~\cite{canf}.}
\label{fig:canf}
\vspace{-0.5cm}
\end{figure}

%\subsection{Augmented Normalizing Flow-based Conditional Inter-frame Coder}
We base our scheme on a conditional augmented normalizing flow-based (CANF-based) inter-frame codec~\cite{canf}. CANF encodes an input frame $x_t$ conditioned on its motion-compensated reference frame $x_c$. Fig.~\ref{fig:canf} depicts the framework of CANF-based inter-frame codec. The encoding process transforms the augmented inputs $(x_t, e_z, e_h)$ into the latent representations $(x_2, \hat{z}_2, \hat{h}_2)$ by conditional autoencoding and hyperprior transforms. The latent variables $\hat{z}_2$ and $\hat{h}_2$ captures the information needed to signal the transformation from the input $x_t$ to $x_2$, which is regularized to approximate $x_c$ during training. The decoding process first sets $x_2$ to $x_c$, followed by decoding $\hat{z}_2$ and $\hat{h}_2$ to perform inverse transformation from $x_c$ to $x_t$.

\begin{figure}
% \vspace{-.6em}
\centering
\includegraphics[width=0.85\linewidth]{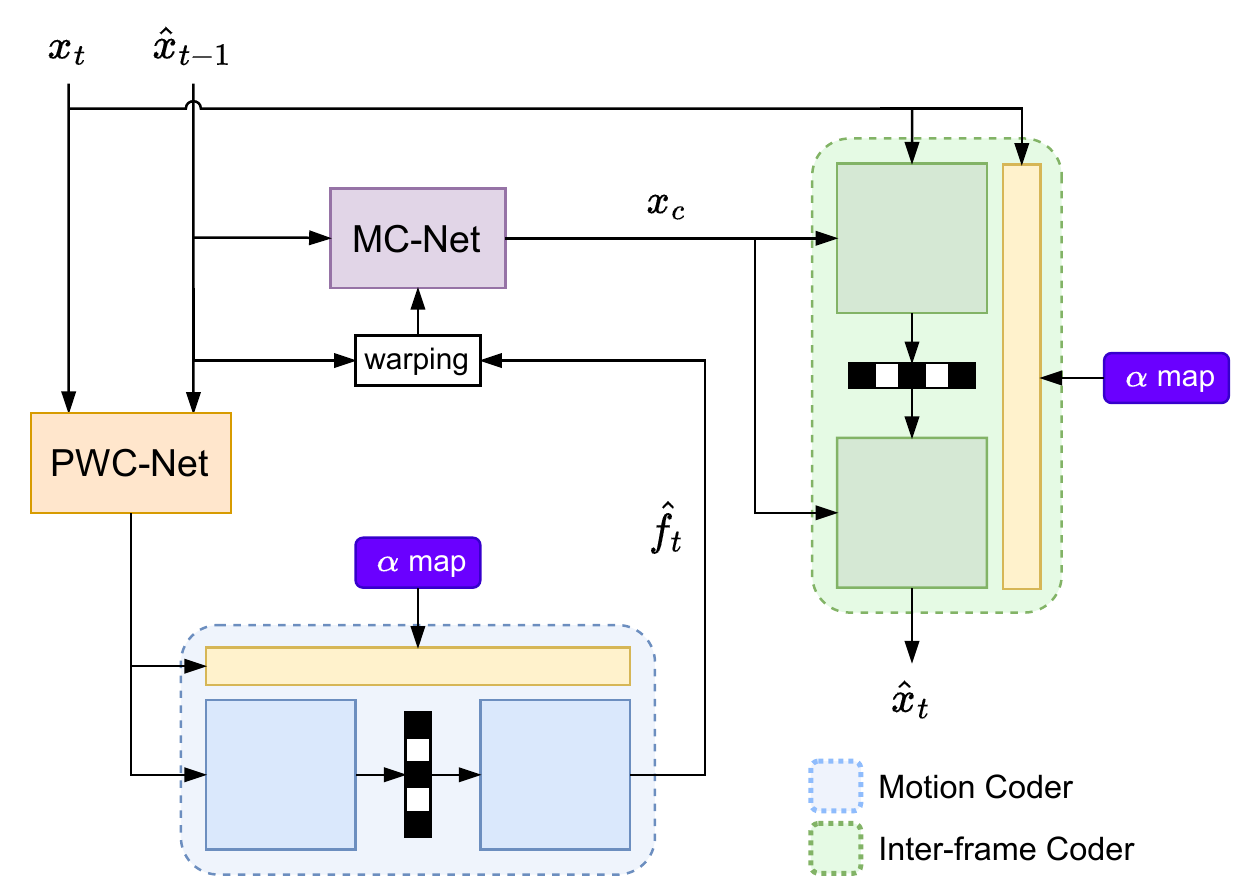}
\caption{The P-frame coding architecture with our proposed scheme.}
\label{fig:arch}
\vspace{-0.5cm}
\end{figure}

% \begin{figure}
% % \vspace{-.6em}
% \centering
% % \includegraphics[width=\linewidth]{figure/ISCAS-architecture-P.png}
% \includegraphics[width=0.8\linewidth]{figures/PCS-SFT_layer.pdf}
% \caption{The detail architecture of the Spatial Feature Transform (SFT) layer, modified from~\cite{sft}.}
% \label{fig:sft}
% \vspace{-0.5cm}
% \end{figure}

\begin{figure*}
% \vspace{-.6em}
\centering
\includegraphics[width=0.77\linewidth]{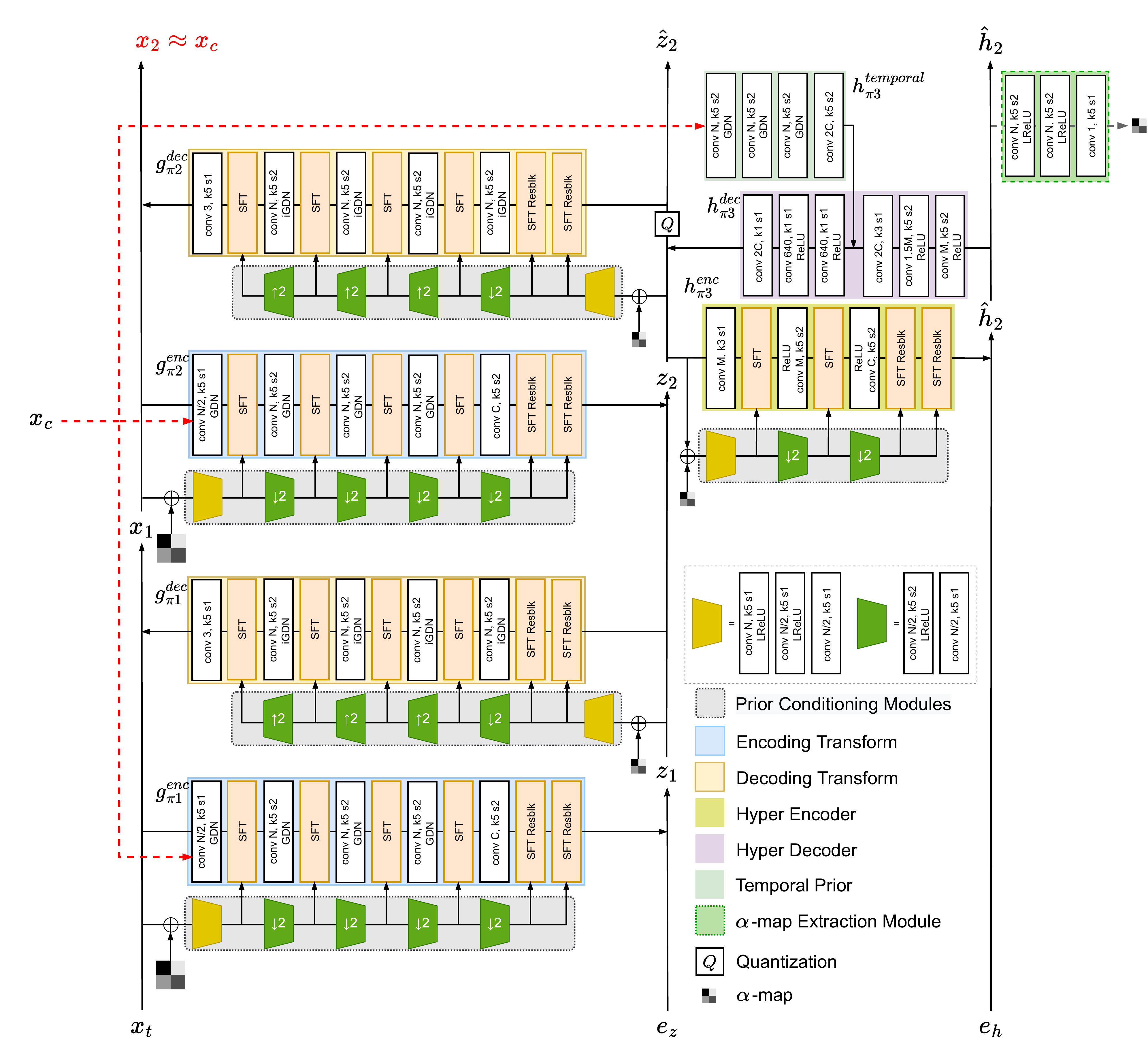}
\caption{The detailed architecture of our \textit{$\alpha$-map}-guided conditional inter-frame codec, where N=C=128, M=192. Following~\cite{canf}, we fix $e_z$ at 0 during training and evaluation. For the hyperprior branch, we draw $e_h \sim \mathcal{U}(-0.5,0.5)$ for simulating the quantization of the hyperprior $\hat{h}_2$ during training, and set it to 0 when $\hat{h}_2$ is rounded during evaluation.}
\label{fig:coder}
\vspace{-0.5cm}
\end{figure*}

\begin{figure}
% \vspace{-.3em}
\centering
\begin{subfigure}{0.3\linewidth}
    % include first image
    \centering
    % \vspace{-1em}
    \includegraphics[width=\linewidth]{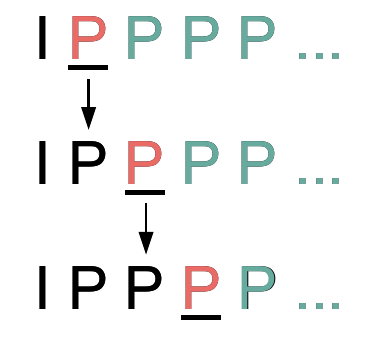}
    \vspace{-1.7em}
    \caption{}
    \vspace{-0.4em}
    \label{fig:alg1}
\end{subfigure}
\begin{subfigure}{0.3\linewidth}
    % include first image
    \centering
    % \vspace{-1em}
    \includegraphics[width=\linewidth]{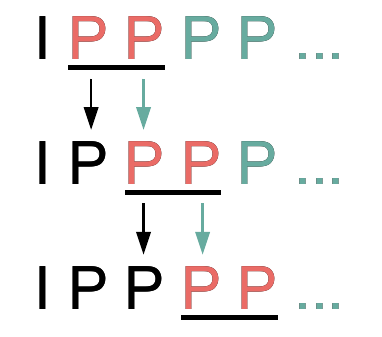}
    \vspace{-1.7em}
    \caption{}
    \vspace{-0.4em}
    \label{fig:alg2}
\end{subfigure}
\caption{The determination of the \textit{$\alpha$ map} (a) without and (b) with the look-ahead mechanism. Black, red, and gray colors refer to coded frames, coding frames, and frames to be coded, respectively.}
\label{fig:alg}
\vspace{-1em}
\end{figure}

\vspace{-0.3em}
\section{Proposed Method}
%In this section, we ...
\subsection{System Overview}
\label{Sec:Overview}
Fig.~\ref{fig:arch} depicts the P-frame coding architecture with our proposed scheme. It comprises a flow estimation network (PWC-Net~\cite{pwc}), a motion compensation network (MC-Net), and two \textit{$\alpha$-map}-guided codecs, which are the motion codec and the conditional inter-frame codec. The former encodes the optical flow map estimated between the coding frame $x_t$ and its reference frame $\hat{x}_{t-1}$, while the latter is adapted from~\cite{canf} to encode the coding frame $x_t$ conditionally on the motion-compensated reference frame $x_c$. $x_t$ and $\hat{x}_{t-1}$ are of size $W \times H$. In this work, the \textit{$\alpha$-map} of dimension $W/64 \times H/64$ serves as a prior conditioning signal used to trade off between the bit rates consumed by the motion and the inter-frame codecs. Each component $\alpha_i \in [-1,1]$ in the \textit{$\alpha$-map} is a real number that corresponds to a distinct $64 \times 64$ patch $i$ in the input frame. By altering the \textit{$\alpha$-map}, a spatially-varying trade-off between the bit rates for motion coding and inter-frame coding is achieved. Moreover, the \textit{$\alpha$-map} is adapted on a frame-by-frame basis, allowing frame-adaptive optimization. 

%The following sections present in detail (1) how the motion and the inter-frame codecs are made conditional on the \textit{$\alpha$-map} and (2) our strategies for determining the \textit{$\alpha$-map}. 

%The coding process begins with the flow estimated model (PWC-Net~\cite{pwc}), which estimates the optical flow between coding frame $x_t$ and reference frame $\hat{x}_{t-1}$, then the estimated flow send to the motion coder with the given \textit{$\alpha$ map}. The reference frame is then warped by the reconstructed flow $\hat{f}_t$, and both the warped frame and original reference frame undergo the MC-Net to form the motion-compensated frame $\Tilde{x}_t$. Finally, the motion-compensated frame $\Tilde{x}_t$ is served as a condition for the conditional inter-frame coder, to reconstruct the final predicted frame $\hat{x}_t$. 

\vspace{-0.4em}
\subsection{Conditional Feature Transformation with The $\alpha$-Map}
% \vspace{-0.4em}
\label{Sec:bft}
%Fig.~\ref{fig:coder} depicts the architecture of the \textit{$\alpha$ map}-guided conditional inter-frame coder. 

%\textcolor{red}{Our base model is an ANF-based coder ANFIC~\cite{anfic}, which aims to transform the input $x$ and one augmented noise $e_z$ into the latent representation $x_2$, $\hat{z}_2$ and $\hat{h}_2$. The latent variables $\hat{z}_2$ and $\hat{h}_2$ are expected to capture most of the information about the input x, therefore force the x2 to approximate 0. Then we only need to transmit $\hat{z}_2$ and $\hat{h}_2$, while $x_2$ is replaced with 0 during inverse transform. Different from ANFIC, which models $p(x)$ unconditionally. The ANF-based inter-frame coder models $p(x_t|x_c)$. The way we condition the $x_c$ is to concatenate $x_c$ with inputs for autoencoding transform. With the help of the conditioning variable $x_c$, our conditional ANF transforms the augmented inputs into $x_c$, $\hat{z}_2$ and $\hat{h}_2$. Similar to ANFIC, we only transmit $\hat{z}_2$ and $\hat{x}_2$, and replace the $x_2$ with $x_c$ during the decoding process.}

To adapt the P-frame coding pipeline to the \textit{$\alpha$-map}, we incorporate Spatial Feature Transform (SFT) layers and SFT Residual Blocks (SFT Resblk)~\cite{qmap, sft} into the motion and the conditional inter-frame codecs. Take our CANF-based inter-frame codec in Fig.~\ref{fig:coder} as an example. SFT applies spatially-adaptive affine transformation to the latent features $F$ in the encoding/decoding transforms, with the element-wise affine parameters ($\gamma, \beta$) derived from the prior conditioning modules. In other words, $SFT(F|\gamma, \beta) = \gamma \odot F + \beta$. It is to be noted that the inputs to the prior conditioning modules include not only the \textit{$\alpha$-map}, but also the corresponding input signals to the encoding/decoding transforms. Before being concatenated with these input signals, the \textit{$\alpha$-map} (of size $W/64 \times H/64$) is scaled accordingly to match their dimensions. 

During the encoding process, the target $x_t$ is transformed into $x_c$, where the latent $\hat{z}_2$ captures the information needed to signal the transformation while the \textit{$\alpha$-map} determined externally (Section~\ref{Sec:ca}) is fed to the prior conditioning modules to adapt the latent features. In particular, we signal the \textit{$\alpha$-map} implicitly in the hyperprior $\hat{h}_2$. That is, during decoding when $x_t$ is recovered from $x_c$, an approximate \textit{$\alpha$-map} is extracted from $\hat{h}_2$ by a lightweight network. Notably, the discrepancy in \textit{$\alpha$-map} for encoding and decoding may contribute to the reconstruction error of $x_t$.

\textcolor{black}{Our motion codec follows a similar architecture to~\cite{anfic} and is likewise guided by the \textit{$\alpha$-map}.}

%For the design of the motion coder, which is similar to the inter-frame coder but without condition input $x_c$ and temporal prior~\cite{dcvc}.

 %During the inverse transform, one simple network then extracts the \textit{$\alpha$ map} from $\hat{h}_2$. 

%Again, one $\alpha$ value corresponds to a 64x64 block of input frames. Therefore, adjusting the $\alpha$ value of \textit{$\alpha$ map} then controls the output rate of the coder. 
\vspace*{-0.5em}
\subsection{Training Objective}
\label{Sec:wl}

We adopt the following objective function to train our system end-to-end. The patch-level bit rate $R_{M_i}$ for motion coding is weighted exponentially with a factor $\delta^{\alpha_i}$ against the patch-level bit rate $R_{R_i}$ for inter-frame coding according to the $\alpha$-map.
\begin{alignat}{2}
\label{equ:rd_loss}
 & L = \lambda \times D+R_W, 
 \\
 & R_W = \sum_{i=1}^N \delta^{\alpha_i} \times R_{M_i}+R_{R_i},
% \vspace{-10em}
\end{alignat}
where the base $\delta=10$ of the exponential is chosen empirically to compensate for the uneven ratio between $R_{M_i}$ and $R_{R_i}$. $N$ is the number of $64 \times 64$ patches in the input frame. It is seen that the model is trained to suppress $R_{M_i}$ for higher $R_{R_i}$ when $\alpha_i = 1$ and otherwise when $\alpha_i = -1$. $R_{M_i},R_{R_i}$ are weighted equally by setting $\alpha_i = 0$. During training, the \textit{$\alpha$-map} is randomized by having $\alpha = tanh(x)$, where $x$ is drawn from a standard normal distribution. This ensures that the model is able to react to any \textit{$\alpha$-map} given at inference time. 

%$applying are randomly generated by \mathcal{N}(0,I)$tanh()$, which the hyperbolic tangent restrict the $\alpha$ values to be within $[-1,1]$. And the zero-mean ensures the original balance between $D$ and $R$.

%By adding the weight term on $R_M$, we are able to adjust the bit-allocation between $R_M$ and $R_R$. For example, the loss leads to suppress the motion rate when $\alpha = 1$ and vice versa. Note that when $\alpha = 0$, the weight loss is equivalent to Eq.\ref{equ:rd_loss_dvc}, which learns the the model learns the average bit-allocation between $R_M$ and $R_R$.
%To train the model that is aware of the \textit{$\alpha$ map}, we proposed the loss function that weighted the rate map with \textit{$\alpha$ map} block-wisely. In typical training loss for learned video compression~\cite{dvclu}: 
%\begin{align}
%\label{equ:rd_loss_dvc}
% L = \lambda \times D+R_M+R_R,
% \vspace{-0.3em}
%\end{align}
%where $\lambda$ is a hyper-parameter to determine the balance between $D$ and $R$. 

%It is however worth noting that the motion rate $R_M$ and the residual rate $R_R$ are weighted equally, and the model learns the average bit-allocation between $R_M$ and $R_R$ across training data. Instead of weighting these rates equally, we proposed to weight $R_M$ and $R_R$ block-wisely by 
% \endgroup

\vspace{-0.4em}
\subsection{Determining The \textit{$\alpha$-Map}}
% \vspace{-0.4em}
\label{Sec:ca}
%In Sec.~\ref{Sec:bft} and Sec.~\ref{Sec:wl}, we discuss how to train a model that can adjust the bit-allocation between the motion rate and the residual rate by given \textit{$\alpha$-map}. However, how to determine the better \textit{$\alpha$-map} in inference time is a critical part of this work. 
We determine the \textit{$\alpha$-map} for content-adaptive bit allocation between motion and inter-frame coding. To this end, we propose an online back-propagation scheme. The idea is to consider the \textit{$\alpha$-map} associated with each input frame as coding parameters to be updated on-the-fly by back-propagation. We take the pre-trained model from Section~\ref{Sec:wl} and minimize Eq.~\eqref{equ:rd_loss} with respect to the \textit{$\alpha$-map}, with $R_W$ taking the form of $\sum_i^N R_{M_i}+R_{R_i}$, where we discard the factor $\delta^{\alpha_i}$ because we wish to arrive at an \textit{$\alpha$-map} that can best trade off between the bit rates for motion and inter-frame coding in order to minimize the rate-distortion cost for the current coding frame. In a sense, this approach is sub-optimal because it optimizes greedily the \textit{$\alpha$-map} of a coding frame without regard to its impacts on future frames (see Fig.~\ref{fig:alg1}).   

To explore the potential of our scheme, we additionally experiment with a look-ahead mechanism that optimizes the \textit{$\alpha$-map} of a coding frame by taking into account its impact on future frames. The idea is illustrated in Fig.~\ref{fig:alg2}, where we minimize the sum of the rate-distortion costs over two consecutive video frames by updating their \textit{$\alpha$-maps} simultaneously. In particular, the resulting \textit{$\alpha$-map} of the first frame in display order is used for coding the first frame, whereas that of the second frame serves as its initial \textit{$\alpha$-map}, which is to be further optimized together with the subsequent frame in a sliding window manner.

\begin{figure*}
\vspace{-.6em}
\centering
\includegraphics[width=0.62\linewidth]{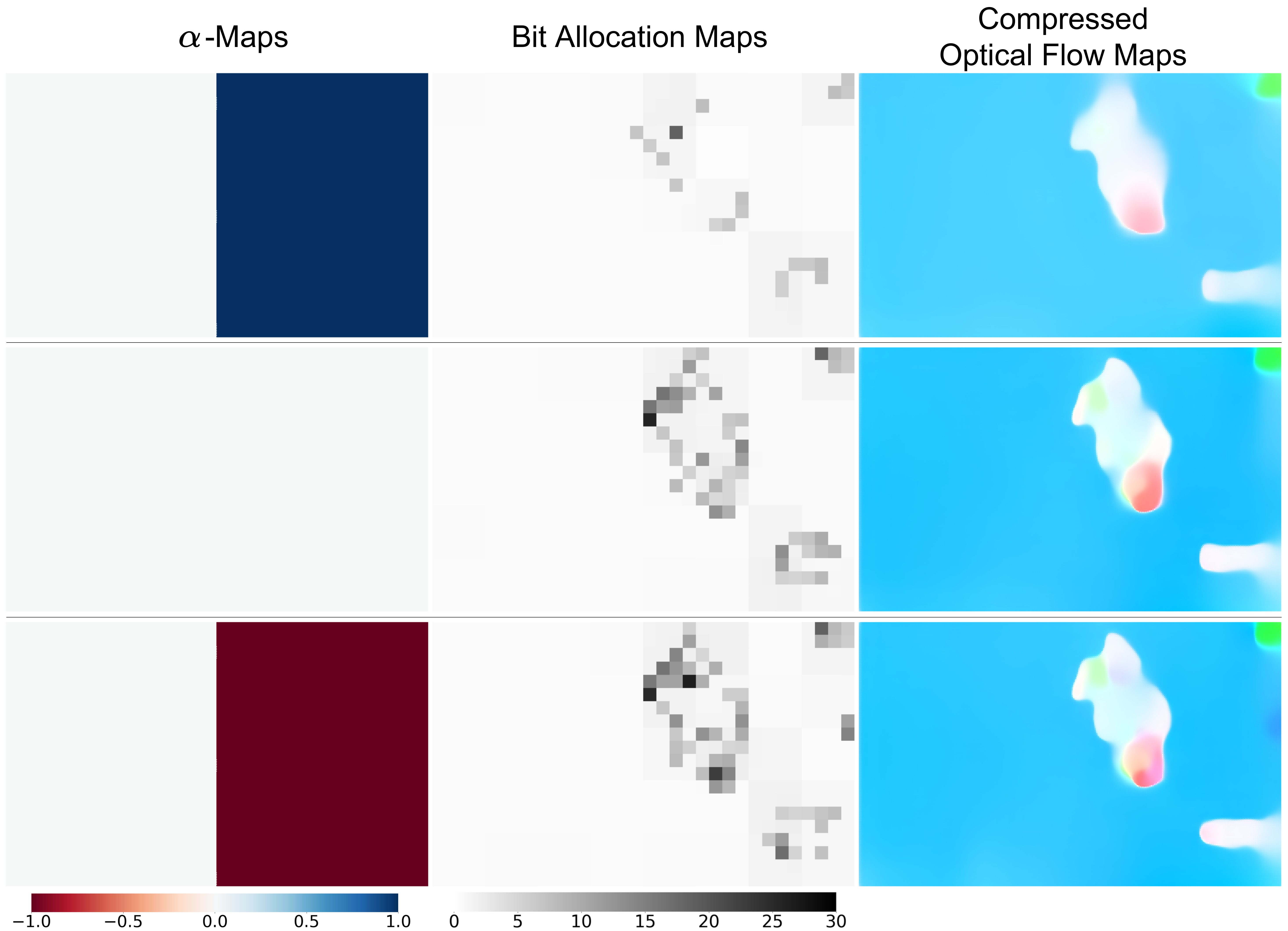}
\caption{Visualization of the \textit{$\alpha$-maps}, the bit allocation maps for motion coding, and the compressed optical flow maps. }
\label{fig:manual_map}
\vspace{-0.5cm}
\end{figure*}
% \vspace{-1em}
\section{Experiments}
%This section details our experimental settings, demonstrates the effectiveness of our \textit{$\alpha$-map} mechanism, shows the rate-distortion performance of the proposed method, and visualizes the optimized \textit{$\alpha$-map}.
\vspace{-.6em}
\subsection{Settings}
{\bf Training details:} 
We use Vimeo-90k dataset~\cite{vimeo} to train our model in two stages. First, we train our model without the prior conditioning modules. We then include the prior conditioning modules to train the entire system end-to-end. The Vimeo-90k dataset contains 91,701 sequences of size $448 \times 256$. Video frames are randomly cropped to $256 \times 256$ for training. We adopt Adam optimizer~\cite{adam}. The learning rate is fixed at $1e^{-4}$ before 300k iterations, and is then decreased to $1e^{-5}$. The $\lambda$ in Eq.~\eqref{equ:rd_loss} is set to 256, 512, 1024, 2048.

{\bf Evaluation methodologies:} 
For evaluation, we test our method on UVG~\cite{uvg}, HEVC Class D~\cite{hevc} and 5 challenging sequences from CLIC'22~\cite{clic22} validation and test datasets. The sequences selected from CLIC'22 have complex motion, e.g. fast motion, zoom-in, or rotation. Remarkably, we downscale the sequences in UVG (of size $1920 \times 1080$) and CLIC'22 (of size $1920 \times 1080$ or $2048 \times 1080$) by a factor of 4 (denoted as UVG$^*$ and CLIC-MIX$^*$ in Table~\ref{tab:exp_BD_PSNR}), in order to facilitate the \textit{$\alpha$-map} optimization with limited GPU memory. We set the intra period to 32, and encode the first 96 frames for all the test sequences. We evaluate the reconstruction quality in PSNR-RGB and the bit-rate in bits-per-pixel (bpp).

{\bf Baseline methods:} The anchor for comparison uses the same learned video codec as ours. Specifically, it adopts ANFIC~\cite{anfic} for I-frame coding and the architecture in Fig.~\ref{fig:arch} for P-frame coding. Particularly, it sets the \textit{$\alpha$-map} uniformly to 0; in other words, there is no content-specific optimization for motion and inter-frame coding. We additionally compare our scheme with  DCVC~\cite{dcvc}, which is the state-of-the-art learned video codec. For a fair comparison, we also use ANFIC~\cite{anfic} as the I-frame codec for DCVC~\cite{dcvc}.\\

\vspace*{.2em}
% Another baseline is \textit{x265} with \textit{veryslow} preset and \textit{low delay} configurations.
% {\bf Experimental Results:}
\begin{table}[]
\centering
\caption{Changes in bit rate for motion and inter-frame coding in response to varying the \textit{$\alpha$-map}.}
\label{tab:rate_div}
\begin{tabular}{lll}
\toprule
\multicolumn{1}{c}{} & \multicolumn{2}{c}{Bit Rate (10$^{-3}$ bpp)} \\ \cline{2-3} 
                     & Motion Coding           & Inter-frame Coding           \\ \hline
$\alpha$=1           & 1.67 (-24.8\%)       & 51.85 (+6.5\%)                                           \\
$\alpha$=0           & 2.22 (0.0\%)                 & 48.67 (0.0\%)                                                    \\
$\alpha$=-1          & 2.86 (+28.8\%)       & 47.13 (-3.2\%)                                           \\ \bottomrule
\end{tabular}
\vspace*{-2em}
\end{table}
\vspace*{-1.8em}
\subsection{Effectiveness of The \textit{$\alpha$-map}}
Table~\ref{tab:rate_div} shows the trade-off between the bit rates used for motion and inter-frame coding on UVG dataset when we alter the \textit{$\alpha$-map} from 1 to -1. In this experiment, the \textit{$\alpha$-map} has a uniform value across spatial locations. It is seen that as compared to $\alpha=0$, the bit rate for motion coding decreases by nearly 25\% (respectively, increases by nearly 29\%) when $\alpha=1$ (respectively, $\alpha=-1$). Accordingly, the \textit{$\alpha$-map} has the opposite effect on the bit rate of inter-frame coding, even though the change is relatively modest.

Fig.~\ref{fig:manual_map} further visualizes how the \textit{$\alpha$-map} impacts the motion bit rate and the quality of the compressed optical flow map patch-wisely. In this experiment, the \textit{$\alpha$-map} is divided into two halves. The left halve has a fixed $\alpha$ value of 0, while that of the right halve changes from 1 to -1 (from top to bottom). We see that the motion bit rate increases with the decreasing $\alpha$ value while the compressed optical flow map has increasing details. These results validate that our model reacts to the given \textit{$\alpha$-map} in the way that we want it to. 
%From the first row to the last row, the compressed optical flow maps are vary from coarse to fine, the bit rate usage of each map also demonstrates the effectiveness of modifying the \textit{$\alpha$-map} patch-wisely.

\subsection{Rate-Distortion Performance}
\vspace*{-.2em}
Table~\ref{tab:exp_BD_PSNR} presents the BD-rate comparison relative to the anchor, which sets the \textit{$\alpha$-map} uniformly to 0. The two variants (Ours$^1$ vs. Ours$^2$) of the proposed method refer to optimizing the \textit{$\alpha$-map} by considering only the current frame and by additionally looking ahead to one future frame, respectively (See Section~\ref{Sec:ca}). From Table~\ref{tab:exp_BD_PSNR}, both variants exhibit 2\%-5\% rate savings than the anchor, and generally outperform DCVC~\cite{dcvc}. The look-ahead variant (Ours$^2$) achieves slightly higher gain at the cost of higher buffering requirements. It is worth noting that the gain of these variants is most obvious on test sequences with fast motion, e.g. Jocky, ReadySteadyGo and RaceHorses. In comparison with the other sequences, they have higher motion bit rates. As such, adapting the motion bit rate to the content of these sequences exerts a more significant influence on the rate-distortion performance. The same observation also holds true for those challenging sequences selected from CLIC'22~\cite{clic22} (CLIC-MIX$^*$). Last but not least, it is seen that conducting content-specific coding optimization through the adaptation of the \textit{$\alpha$-map} consistently shows gain over different test sequences. This suggests that there exists a domain gap between training and individual sequences, and that our scheme is able to help reduce the gap. 

%This phenomenon is reasonable since the bit rates of the motion coder occupied much less then the inter-frame coder for the overall bit rates, the rate deviation then become critical on motion coder. To verify this, we additionally choose 5 special sequences in CLIC'22~\cite{clic22} (CLIC-MIX$^*$).

%higher rate reductions of 4.30\%, 3.89\%, and 3.39\% can be reach in our methods. However, for the static sequences (HoneyBee and ShakeNDry), lower gain have absorbed. 

%On CLIC-MIX$^*$, Seq.\textit{A} is a sequence with zoom-in and considerable occlusion effects. Our method shows 2.04\% rate reduction. Seq.\textit{B} has extremely fast motion, and our method shows 1.59\% rate reduction. Seq.\textit{C} displays a rotation ring in the middle, a significant improvement of 4.49\% rate reduction is observed. Seq.\textit{D} shows the zoom-out and shaking effect, and our method shows 3.89\% rate reduction. Seq.\textit{E} demonstrates an irregular chattering camera with a dancing person, 2.46\% rate reduction is observed in our method.

% {\bf FFNet (optional)}

\begin{figure*}
% \vspace{-.6em}
\vspace{-1cm}
\centering
\includegraphics[width=0.75\linewidth]{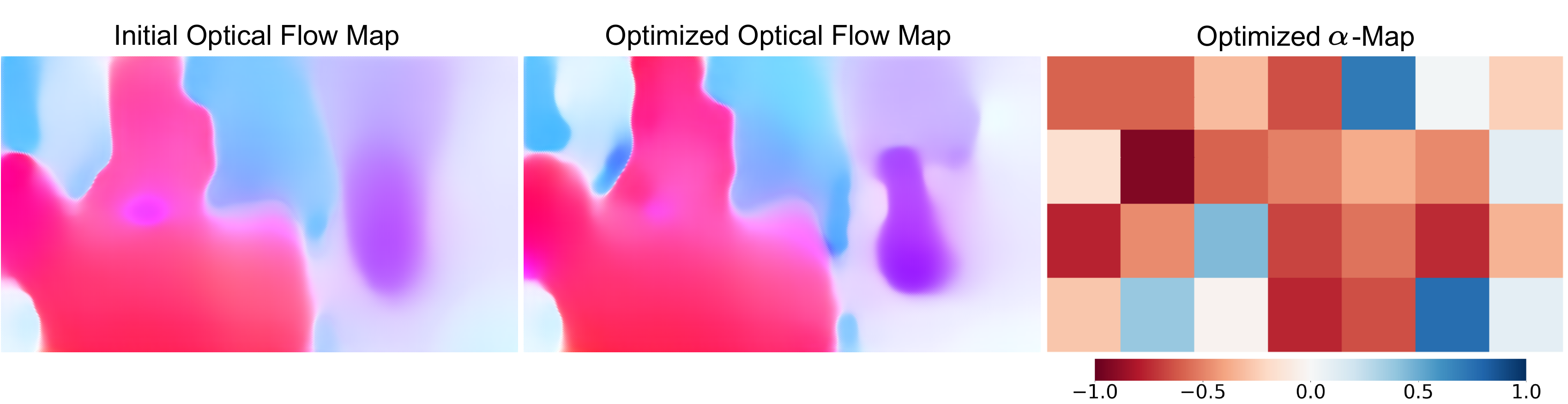}
\vspace{-0.1cm}
\caption{Visualization of the initial optical flow map with $\alpha=0$ (left) and the resulting optical flow map (middle) with the optimized \textit{$\alpha$-map} (right).}
\label{fig:optim_map}
\vspace{-0.5cm}
\end{figure*}

Fig.~\ref{fig:optim_map} visualizes how the compressed optical flow map changes with the \textit{$\alpha$-map} optimization. In this example, the \textit{$\alpha$-map} has a tendency to be decreased in exchange for a higher motion rate to represent the complex structure inherent in the optical flow map. It seen that the resulting flow map preserves more details than the initial flow map compressed with $\alpha=0$.

\begin{table}[]
\centering
\caption{BD-rate comparison.}
\vspace{-0.7em}
\label{tab:exp_BD_PSNR}
\begin{tabular}{ccccc}
\toprule
\multirow{2}{*}{\textbf{Datasets}}     & \multirow{2}{*}{\textbf{ID}} & \multicolumn{3}{c}{\textbf{BD-rate (\%) PSNR}}        \\ \cline{3-5} 
                                       &                              & \textbf{DCVC} & \textbf{Ours$^1$} & \textbf{Ours$^2$} \\ \hline
\multirow{7}{*}{\textbf{UVG$^*$}}      & \textbf{Beauty}              & -0.88         & -2.49             & -2.77             \\
                                       & \textbf{Bosphorus}           & -1.01         & -2.12             & -2.60             \\
                                       & \textbf{HoneyBee}            & 9.42          & -1.36             & -1.51             \\
                                       & \textbf{Jockey}              & -3.19         & -4.30             & -4.61             \\
                                       & \textbf{ReadySteadyGo}       & -8.91         & -3.89             & -4.07             \\
                                       & \textbf{ShakeNDry}           & -15.83        & -0.68             & -1.40             \\
                                       & \textbf{YachtRide}           & 4.58          & -3.12             & -3.28             \\ \hline
\multirow{4}{*}{\textbf{HEVC-D}}       & \textbf{BasketballPass}      & -0.70         & -2.16             & -3.17             \\
                                       & \textbf{BlowingBubbles}      & 5.49          & -3.27             & -3.19             \\
                                       & \textbf{BQSquare}            & 38.91         & -2.47             & -2.66             \\
                                       & \textbf{RaceHorses}          & 5.92          & -3.39             & -3.32             \\ \hline
\multirow{5}{*}{\textbf{CLIC-MIX$^*$}} & \textbf{5d8f0 (zoom-in)}     & 44.32         & -2.04             & -2.28             \\
                                       & \textbf{25f0c (fast motion)} & 7.78          & -1.59             & -2.11             \\
                                       & \textbf{4761c (rotation)}    & 26.41         & -4.49             & -5.16             \\
                                       & \textbf{a89f6 (zoom-out)}    & -4.09         & -3.89             & -3.98             \\
                                       & \textbf{cda52 (shaking)}     & 4.46          & -2.46             & -2.78             \\ \bottomrule
\end{tabular}
\vspace{-1.7em}
\end{table}

% \input{chapter/4-2-2_bd_rate_noDCVC.tex}
% \vspace{-0.8em}
\section{Conclusion}
\vspace{-0.4em}
This paper presents a content-adaptive motion rate adaptation scheme for learned video compression. It uses a bit allocation map to trade off between the bit rates for motion and inter-frame coding in a spatially-adaptive manner. The map is optimized through online back-propagation. Our major findings include: (1) content-adaptive motion rate adaptation helps mitigate the domain gap between training and test data; (2) the proposed scheme is found to be more effective on test sequences with complicated motion; and (3) our look-ahead mechanism is able to better optimize the bit allocation map for higher coding gain.

\vspace{-1em}
 
%%%%%%%%% REFERENCES
{\small
\bibliographystyle{IEEEtran}
\bibliography{egbib}

% Generated by IEEEtran.bst, version: 1.14 (2015/08/26)
\begin{thebibliography}{10}
\providecommand{\url}[1]{#1}
\csname url@samestyle\endcsname
\providecommand{\newblock}{\relax}
\providecommand{\bibinfo}[2]{#2}
\providecommand{\BIBentrySTDinterwordspacing}{\spaceskip=0pt\relax}
\providecommand{\BIBentryALTinterwordstretchfactor}{4}
\providecommand{\BIBentryALTinterwordspacing}{\spaceskip=\fontdimen2\font plus
\BIBentryALTinterwordstretchfactor\fontdimen3\font minus
  \fontdimen4\font\relax}
\providecommand{\BIBforeignlanguage}[2]{{%
\expandafter\ifx\csname l@#1\endcsname\relax
\typeout{** WARNING: IEEEtran.bst: No hyphenation pattern has been}%
\typeout{** loaded for the language `#1'. Using the pattern for}%
\typeout{** the default language instead.}%
\else
\language=\csname l@#1\endcsname
\fi
#2}}
\providecommand{\BIBdecl}{\relax}
\BIBdecl

\bibitem{dvclu}
G.~Lu, W.~Ouyang, D.~Xu, X.~Zhang, C.~Cai, and Z.~Gao, ``Dvc: An end-to-end
  deep video compression framework,'' in \emph{Proceedings of the IEEE/CVF
  Conference on Computer Vision and Pattern Recognition}, 2019, pp.
  11\,006--11\,015.

\bibitem{dvcpro}
G.~Lu, X.~Zhang, W.~Ouyang, L.~Chen, Z.~Gao, and D.~Xu, ``An end-to-end
  learning framework for video compression,'' \emph{IEEE transactions on
  Pattern Analysis and Machine Intelligence}, 2020.

\bibitem{mlvc}
J.~Lin, D.~Liu, H.~Li, and F.~Wu, ``M-lvc: multiple frames prediction for
  learned video compression,'' in \emph{Proceedings of the IEEE/CVF Conference
  on Computer Vision and Pattern Recognition}, 2020, pp. 3546--3554.

\bibitem{elfvc}
O.~Rippel, A.~G. Anderson, K.~Tatwawadi, S.~Nair, C.~Lytle, and L.~Bourdev,
  ``Elf-vc: Efficient learned flexible-rate video coding,'' in
  \emph{Proceedings of the IEEE/CVF International Conference on Computer Vision
  (ICCV)}, October 2021, pp. 14\,479--14\,488.

\bibitem{ssf}
E.~Agustsson, D.~Minnen, N.~Johnston, J.~Balle, S.~J. Hwang, and G.~Toderici,
  ``Scale-space flow for end-to-end optimized video compression,'' in
  \emph{Proceedings of the IEEE/CVF Conference on Computer Vision and Pattern
  Recognition}, 2020, pp. 8503--8512.

\bibitem{fvc}
Z.~Hu, G.~Lu, and D.~Xu, ``Fvc: A new framework towards deep video compression
  in feature space,'' in \emph{Proceedings of the IEEE/CVF Conference on
  Computer Vision and Pattern Recognition}, 2021, pp. 1502--1511.

\bibitem{mmsp}
T.~Ladune, P.~Philippe, W.~Hamidouche, L.~Zhang, and O.~D{\'e}forges, ``Optical
  flow and mode selection for learning-based video coding,'' in \emph{2020 IEEE
  22nd International Workshop on Multimedia Signal Processing (MMSP)}.\hskip
  1em plus 0.5em minus 0.4em\relax IEEE, 2020, pp. 1--6.

\bibitem{dcvc}
J.~Li, B.~Li, and Y.~Lu, ``Deep contextual video compression,'' \emph{Advances
  in Neural Information Processing Systems}, vol.~34, 2021.

\bibitem{canf}
Y.-H. Ho, C.-P. Chang, P.-Y. Chen, A.~Gnutti, and W.-H. Peng, ``Canf-vc:
  Conditional augmented normalizing flows for video compression,'' in
  \emph{European Conference on Computer Vision}, 2022.

\bibitem{campos2019content}
J.~Campos, S.~Meierhans, A.~Djelouah, and C.~Schroers, ``Content adaptive
  optimization for neural image compression,'' in \emph{Proceedings of the
  IEEE/CVF Conference on Computer Vision and Pattern Recognition (CVPR)
  Workshops}, 2019.

\bibitem{zou20202}
N.~Zou, H.~Zhang, F.~Cricri, H.~R. Tavakoli, J.~Lainema, M.~Hannuksela,
  E.~Aksu, and E.~Rahtu, ``L 2 c--learning to learn to compress,'' in
  \emph{2020 IEEE 22nd International Workshop on Multimedia Signal Processing
  (MMSP)}.\hskip 1em plus 0.5em minus 0.4em\relax IEEE, 2020, pp. 1--6.

\bibitem{lu2020content}
G.~Lu, C.~Cai, X.~Zhang, L.~Chen, W.~Ouyang, D.~Xu, and Z.~Gao, ``Content
  adaptive and error propagation aware deep video compression,'' in
  \emph{European Conference on Computer Vision}.\hskip 1em plus 0.5em minus
  0.4em\relax Springer, 2020, pp. 456--472.

\bibitem{van2021instance}
T.~van Rozendaal, J.~Brehmer, Y.~Zhang, R.~Pourreza, and T.~S. Cohen,
  ``Instance-adaptive video compression: Improving neural codecs by training on
  the test set,'' \emph{arXiv preprint arXiv:2111.10302}, 2021.

\bibitem{zhang2021learn}
H.~Zhang, F.~Cricri, H.~R. Tavakoli, M.~Santamaria, Y.-H. Lam, and M.~M.
  Hannuksela, ``Learn to overfit better: finding the important parameters for
  learned image compression,'' in \emph{2021 International Conference on Visual
  Communications and Image Processing (VCIP)}.\hskip 1em plus 0.5em minus
  0.4em\relax IEEE, 2021, pp. 1--5.

\bibitem{brand2021rate}
F.~Brand, K.~Fischer, and A.~Kaup, ``Rate-distortion optimized learning-based
  image compression using an adaptive hierachical autoencoder with conditional
  hyperprior,'' in \emph{Proceedings of the IEEE/CVF Conference on Computer
  Vision and Pattern Recognition (CVPR) Workshops}, 2021, pp. 1885--1889.

\bibitem{qmap}
M.~Song, J.~Choi, and B.~Han, ``Variable-rate deep image compression through
  spatially-adaptive feature transform,'' in \emph{Proceedings of the IEEE/CVF
  International Conference on Computer Vision}, 2021, pp. 2380--2389.

\bibitem{hu2022coarse}
Z.~Hu, G.~Lu, J.~Guo, S.~Liu, W.~Jiang, and D.~Xu, ``Coarse-to-fine deep video
  coding with hyperprior-guided mode prediction,'' in \emph{Proceedings of the
  IEEE/CVF Conference on Computer Vision and Pattern Recognition}, 2022, pp.
  5921--5930.

\bibitem{pwc}
D.~Sun, X.~Yang, M.-Y. Liu, and J.~Kautz, ``Pwc-net: Cnns for optical flow
  using pyramid, warping, and cost volume,'' in \emph{Proceedings of the IEEE
  conference on computer vision and pattern recognition}, 2018, pp. 8934--8943.

\bibitem{sft}
X.~Wang, K.~Yu, C.~Dong, and C.~C. Loy, ``Recovering realistic texture in image
  super-resolution by deep spatial feature transform,'' in \emph{Proceedings of
  the IEEE conference on computer vision and pattern recognition}, 2018, pp.
  606--615.

\bibitem{anfic}
Y.-H. Ho, C.-C. Chan, W.-H. Peng, H.-M. Hang, and M.~Domański, ``Anfic: Image
  compression using augmented normalizing flows,'' \emph{IEEE Open Journal of
  Circuits and Systems}, vol.~2, pp. 613--626, 2021.

\bibitem{vimeo}
T.~Xue, B.~Chen, J.~Wu, D.~Wei, and W.~T. Freeman, ``Video enhancement with
  task-oriented flow,'' \emph{International Journal of Computer Vision}, vol.
  127, no.~8, pp. 1106--1125, 2019.

\bibitem{adam}
J.~B. Diederik P.~Kingma, ``Adam: A method for stochastic optimization,''
  \emph{International Conference for Learning Representations}, 2015.

\bibitem{uvg}
A.~Mercat, M.~Viitanen, and J.~Vanne, ``Uvg dataset: 50/120fps 4k sequences for
  video codec analysis and development,'' in \emph{Proceedings of the 11th ACM
  Multimedia Systems Conference}, 2020, pp. 297--302.

\bibitem{hevc}
G.~J. Sullivan, J.-R. Ohm, W.-J. Han, and T.~Wiegand, ``Overview of the high
  efficiency video coding (hevc) standard,'' \emph{IEEE Transactions on
  circuits and systems for video technology}, vol.~22, no.~12, pp. 1649--1668,
  2012.

\bibitem{clic22}
``"5th challenge on learned image compression",'' \emph{URL
  http://compression.cc}, 2022.

\end{thebibliography}
}

\end{document}